\documentstyle[mprocl]{article}

\def\be{\begin{equation}}
\def\ee{\end{equation}}
\def\bea{\begin{eqnarray}}
\def\eea{\end{eqnarray}}

\begin{document}

\title{GRAVITATIONAL MICROLENSING RESEARCH}

\author{PHILIPPE JETZER}
\address{Paul Scherrer Institute, Laboratory for Astrophysics, CH-5232 
Villigen PSI, and
Institute of Theoretical Physics, University of Z\"urich, 
Winterthurerstrasse
190, CH-8057 Z\"urich, Switzerland}

\maketitle
\abstracts{ 
One of the most important problems in astrophysics concerns the nature
of the dark matter in galactic halos, whose presence is 
implied mainly by the 
observed flat rotation curves in spiral galaxies.
In the framework of a baryonic scenario the most plausible 
candidates are brown dwarfs, M-dwarfs 
or white dwarfs and cold molecular clouds
(mainly of $H_2$).
The former can be detected with the ongoing microlensing experiments,
which are rapidly leading to important new results.
The French collaboration EROS and the American-Australian
collaboration MACHO have reported until August 1997 the observation of 
$\sim$ 16 microlensing
events by monitoring during several years the brightness of millions
of stars in the Large Magellanic Cloud and one event towards
the Small Magellanic Cloud. 
In particular, the MACHO team found 
8 microlensing candidates by analysing their first 2 years
of observations. This implies that the halo dark matter
fraction in form of MACHOs 
(Massive Astrophysical Compact Halo Objects)
is of the order of 45-50\% assuming a standard spherical halo model.
More than 150 microlensing events have been detected in the direction
of the galactic bulge by the MACHO, OGLE and DUO teams.
The measured optical depth implies the presence of a bar in the
galactic centre. 
Here, we give an overview of microlensing and the main
results achieved so far.}


\section{Introduction}

One of the most important problems in astrophysics concerns
the nature of the dark matter in galactic halos, whose presence is 
implied by the observed flat rotation curves in spiral galaxies
\cite{kn:Faber,kn:Trimble}, the X-ray diffuse emission
in elliptical galaxies as well as by the dynamics of galaxy
clusters. Primordial nucleosynthesis entails that most of the baryonic 
matter in the Universe is nonluminous, and such an amount of dark matter falls
suspiciously close to that required by the rotation curves.
Surely, the standard model of elementary particle forces can hardly
be viewed as the ultimate theory and all the attempts in that
direction invariably call for new particles. Hence, the idea of
nonbaryonic dark matter naturally enters the realm of cosmolgy
and may help in the understanding of the process of galaxy
formation and clustering of galaxies.  

The problem of dark matter started already with the pioneering work
of Oort \cite{kn:Oort} in 1932 and Zwicky \cite{kn:Zwicky} in 1933
and its mistery is still not solved.
Actually, there are several dark matter problems on different scales
ranging from the solar neighbourhood, galactic halos, cluster of galaxies
to cosmological scales. Dark matter is also needed to understand
the formation of large scale structures in the universe.
Many candidates have been proposed, either baryonic or not,
to explain dark matter.

Here, we discuss the dark matter problem in the halo of our
Galaxy in connection with 
microlensing searches. We present the basics of microlensing and
an overview of the results obtained so far, without,
however, being exhaustive and this also for
the references we quote. 
The content is as follows:
first, we review the evidence for dark matter 
in the halo of our Galaxy. 
In Section 3 we present the baryonic candidates for dark matter
and in Section 4 we give an overview of the results of
microlensing searches achieved so far. 
In Section 5 we discuss the basics of microlensing
(optical depth, microlensing rates, etc.) and
in Section 6 we briefly present 
a scenario in which part of the dark matter
is in the form of cold molecular clouds (mainly of $H_2$).

\section{Mass of the Milky Way} 

The best evidence for dark matter in galaxies comes from the rotation
curves of spirals.
Measurements of the rotation velocity $v_{rot}$ of stars up to the 
visible edge 
of the spiral galaxies and of $HI$ gas in the disk beyond the
optical radius (by measuring the 
Doppler shift in the 21-cm line) imply that $v_{rot} \approx$ constant
out to very large distances, rather than to show a Keplerian falloff.
These observations started around 1970 \cite{kn:Rubin}, thanks
to the improved sensitivity in both optical and 21-cm bands.
By now there are observations for over thousand 
spiral galaxies with reliable
rotation curves out to large radii \cite{kn:Persic}. In almost all of them
the rotation curve is flat or slowly rising out to the last measured
point. Very few galaxies show falling rotation curves and those
that do either fall less rapidly than Keplerian have nearby companions
that may perturb the velocity field or have large spheroids
that may increase the rotation velocity near the centre.

There are also measurements of the rotation velocity for our Galaxy.
However, these observations turn out to be rather difficult, and
the rotation curve has been measured only up to a distance of about
20 kpc. Without any doubt our own galaxy has a typical flat 
rotation curve.
A fact this, which implies 
that it is possible to search directly for dark matter
characteristic of spiral galaxies in our own Milky Way.

In oder to infer the total mass one can also study the proper
motion of the Magellanic Clouds and of other satellites of our
Galaxy.
Recent studies \cite{kn:Zaritsky,kn:Lin,kn:Kochanek}
do not yet allow an accurate 
determination of $v_{rot}(LMC)/v_0$ 
($v_0 = 210 \pm 10$ km/s  being the local rotational velocity).
Lin et al. \cite{kn:Lin}  
analyzed the proper motion observations and concluded that 
within 100 kpc the Galactic halo has a mass 
$\sim 5.5 \pm 1 \times 10^{11} M_{\odot}$ and a substantial fraction 
$\sim 50\%$ of this mass is distributed beyond the present distance
of the Magellanic Clouds of about 50 kpc. Beyond 100 kpc the mass may 
continue to increase to $\sim 10^{12} M_{\odot}$ within its tidal radius
of about 300 kpc. This value for the total mass of the Galaxy is in
agreement with the results of Zaritsky et al. \cite{kn:Zaritsky}, who found
a total mass in the range 9.3 to 12.5 $\times 10^{11} M_{\odot}$, the former
value by assuming radial satellite orbits whereas the latter by assuming
isotropic satellite orbits.
 
The results of Lin et al. \cite{kn:Lin} suggest that
the mass of the halo dark matter up to the Large Magellanic Cloud
(LMC) is roughly half of the value
one gets for the standard halo model (with flat rotation
curve up to the LMC and spherical shape), implying thus the same reduction
for the number of expected microlensing events.
Kochanek \cite{kn:Kochanek} analysed the global mass distribution of the
Galaxy adopting a Jaffe model, whose parameters are determined
using the observations on the
proper motion of the satellites of the Galaxy, the Local
Group timing constraint and the ellipticity of the M31 orbit. 
With these observations Kochanek \cite{kn:Kochanek}
concludes that the mass inside 50 kpc is $5.4 \pm 1.3 \times 
10^{11} M_{\odot}$.
This value becomes, however, slightly smaller when using only the satellite 
observations and the disk rotation constraint, in this case
the median mass interior to 50 kpc is in the interval 3.3 to 6.1
(4.2 to 6.8) without (with) Leo I satellite in units of $10^{11} M_{\odot}$.
The lower bound without Leo I is 65\% of the mass expected assuming
a flat rotation curve up to the LMC.

\section{Baryonic dark matter candidates}

Before discussing the baryonic dark matter
we would like to mention that another
class of candidates which is seriously taken into consideration
is the so-called cold dark matter, which
consists for instance of axions
or supersymmetric particles like neutralinos \cite{kn:jungman}.
Here, we will not discuss cold dark matter
in detail. However, recent studies
seem to point out that there is a discrepancy between the calculated (through
N-body simulations)
rotation curve for dwarf galaxies assuming an halo of cold dark matter
and the measured curves \cite{kn:moore,kn:navarro,kn:Silk}. 
If this fact is confirmed, this
would exclude cold dark matter as a major constituent of the 
halo of dwarf galaxies and possibly also of spiral
galaxies.

From the Big Bang nucleosynthesis model \cite{kn:copi,kn:PDG} 
and from the observed 
abundances of primordial elements one infers:
$0.010 \leq h^2_0 \Omega_B \leq 0.016$ or
with $h_0 \simeq 0.4 - 1$ one gets $0.01 \leq \Omega_B \leq 0.10$
(where $\Omega_B = \rho_B /\rho_{crit}$, and $\rho_{crit}=3H_0^2/8\pi G$).
Since for the amount of luminous baryons one finds
$\Omega_{lum} \ll \Omega_B$, it follows that an
important fraction 
of the baryons are dark.
In fact the dark baryons may well make up the entire dark halo matter.

The halo dark matter cannot be in the form of hot 
ionized hydrogen gas otherwise there would be a large
X-ray flux, for which there are stringent upper limits \cite{kn:corx}.
The abundance of neutral hydrogen gas
is inferred from the 21-cm measurements, which show that its contribution is 
small. Another possibility is that the hydrogen gas is in molecular form
clumped into cold clouds, as  
we will briefly discuss in Section 6.
Baryons could otherwise have been processed in stellar remnants
(for a detailed discussion see \cite{kn:Carr}).
If their mass is below $\sim0.08~M_{\odot}$ they are too light to ignite
hydrogen burning reactions. 
The possible origin of such brown dwarfs or Jupiter like bodies
(called also MACHOs),
by fragmentation or by some other mechanism, is at present
not well understood.  It has also been pointed out that the mass distribution
of the MACHOs, normalized to the dark halo mass density, could be 
a smooth continuation of the known initial mass function 
of ordinary stars
\cite{kn:Derujula1}. 
The ambient radiation, or their own body heat, would make
sufficiently small objects of H and He evaporate rapidly.
The condition that the rate of evaporation of such a hydrogenoid sphere be
insufficient to halve its mass in a billion years leads to the 
following lower limit on their mass \cite{kn:Derujula1}: 
$M > 10^{-7} M_{\odot}(T_S /30~ K)^{3/2} (1~ g~cm^{-3}/ \rho)^{1/2}$
($T_S$ being their surface
temperature and $\rho$ their average density, which we expect
to be of the order $\sim 1~ g~ cm^{-3}$).

Otherwise, 
MACHOs might be either M-dwarfs or else white dwarfs.
As a matter of fact, a deeper analysis shows that the M-dwarf option
looks problematic. The null result of several searches for low-mass stars
both in the disk and in the halo of our
Galaxy suggests that the halo cannot be mostly in the form of hydrogen
burning main sequence M-dwarfs. Optical imaging of high-latitude
fields taken with the Wide Field Camera of the Hubble Space Telescope
indicates that less than $\sim 6\%$ of the halo can be in this
form \cite{kn:JBahcall}. 
Observe, however, that these results are derived under the assumption
of a smooth spatial distribution of M-dwarfs, and become considerably
less severe in the case of a clumpy distribution \cite{kn:Kerins,kn:Kerins1}.

A scenario
with white dwarfs as a major constituent of the galactic halo
dark matter has been explored \cite{kn:Tamanaha}.
However, it requires a rather ad hoc initial mass function sharply 
peaked around 2 - 6 $M_{\odot}$. Future
Hubble deep field exposures could either find the white dwarfs 
or put constraints on their fraction in the halo \cite{kn:Kawaler}.
Also a substantial component of neutron
stars and black holes with mass higher than $\sim 1~M_{\odot}$ 
is excluded, for otherwise they would  lead to an overproduction of heavy 
elements relative to the observed abundances. 

\section{Present status of microlensing research}

It has been pointed out by Paczy\'nski \cite{kn:Paczynski} that microlensing 
allows the detection of MACHOs located in the galactic halo in the mass
range \cite{kn:Derujula1}  
$10^{-7} < M/M_{\odot} <  1$, as well as MACHOs in the disk
or bulge of our Galaxy \cite{kn:Paczynski1991,kn:Griest2}.
Since this first proposal microlensing searches have turned very quickly
into reality and in about a decade they have become an important
tool for astrophysical investigations. 
Microlensing is very promising for the search of
planets around other stars in our Galaxy and generates also 
very large databases for variable stars, a field which has already
benefitted a lot. Because of to the
increase of observations, due also to the fact
that new experiments are becoming operative, the situation is changing
rapidly and, therefore,
the present results should be considered as preliminary.
Within few years the amount of data will be such that several problems
will be solved or at least allow to achieve substantial progress.
The following
presentation is also not exhaustive with respect to
all what has been found so far.

\subsection{Towards the LMC and the SMC}

In September 1993 the French collaboration EROS \cite{kn:Aubourg}
announced the discovery of 2 microlensing candidates
and the American--Australian
collaboration MACHO of one candidate \cite{kn:Alcock}
by monitoring stars in the LMC.

In the meantime the MACHO team reported the observation of
altogether 8 events
(one is a binary lensing event) analysing
their first two years of data
by monitoring about 8.5 million 
of stars in the LMC \cite{kn:Pratt}. 
The inferred optical depth is $\tau = 2.1^{+1.1}_{-0.7} \times 10^{-7}$
when considering 6 events \footnote{In fact, the two disregarded events are 
a binary lensing and one which is rated as marginal.} (see Table 3)
(or $\tau = 2.9^{+1.4}_{-0.9} \times 10^{-7}$ when
considering all the 8 detected events). Correspondingly, this implies
that about 45\% (50\% respectively) of the halo dark matter is in form of 
MACHOs and they find an average mass $0.5^{+0.3}_{-0.2} M_{\odot}$ 
assuming a standard spherical halo model.
It may well be that there is also a contribution of events due
to MACHOs located in the LMC itself or in a thich disk of our galaxy,
the corresponding optical depth is estimated to be only \cite{kn:Pratt}
$\tau=5.4 \times 10^{-8}$.
Other events have been detected towards the LMC by the MACHO group, which 
have been put on their list of alert events. In particular 2 events are
reported in 1996 and already 4 events in 1997. The full analysis of
the 1996 and 1997 seasons is still not published.

EROS has also searched for very-low mass MACHOs by looking for
microlensing events with time scales ranging from 30 minutes to 
7 days \cite{kn:EROS}. The lack of candidates in this range 
places significant constraints on any model for the halo that relies
on objects in the range $5 \times 10^{-8} < M/M_{\odot} < 2 \times 10^{-2}$.
Indeed, such objects may make up at most 20\% of the halo dark matter
(in the range between $5 \times 10^{-7} < M/M_{\odot} < 2 \times
10^{-3}$ at most 10\%). 
Similar conclusions have also been reached by the MACHO group
\cite{kn:Pratt}. 
Recently, the MACHO team reported \cite{kn:Alcock2} the first discovery of a 
microlensing event towards the Small Mgellanic Cloud (SMC). The full
analysis of the four years 
data on the SMC is still underway, so that more candidates
may be found in the near future. A rough estimate 
of the optical depth leads to about the same value as found
towards the LMC.

The EROS group has completed his first campaign based on photographic
plates and CCD, the latter one for the short duration events.
Since the middle of 1996 the EROS group has put into operation a new
1 meter telescope, located in La Silla (Chile), and which is fully
dedicated to microlensing searches using CCD cameras. The improved experiment
is called EROS II.

\subsection{Towards the galactic centre}

Towards the galactic bulge the 
Polish-American team OGLE \cite{kn:Udalski}
announced his first event also in September 1993.
Since then OGLE found in their data from the 1992 - 1995 observing
seasons altogether 18 microlensing events (one being a binary lens).
Based on their first 9 events the OGLE team estimated the optical depth
towards the bulge as \cite{kn:udal}  $\tau = (3.3 \pm 1.2) \times 10^{-6}$.
This has to be compared with the theoretical calculations which
lead to a value \cite{kn:Paczynski1991,kn:Griest2} 
$\tau \simeq (1 - 1.5)\times 10^{-6}$, which does, however,
not take into account the contribution of lenses in the bulge itself,
which might well explain the discrepancy. In fact, when taking into
account also the effect of microlensing by galactic bulge stars
the optical depth gets bigger \cite{kn:Kiraga}
and might easily be compatible with the measured value.
This implies the presence of a bar in the galactic centre.  
In the meantime the OGLE group got a new dedicated 1.3 meter
telescope located at the Las Campanas Observatory. The OGLE-2 collaboration
has started the observations in 1996 and is monitoring the
bulge, the LMC and the SMC as well.

The French DUO \cite{kn:Alard} team found 12 microlensing events (one of
which being a binary event) by monitoring the galactic bulge during the 1994
season with the ESO 1 meter Schmidt telescope. The photographic plates were
taken in two different colors to test achromaticity.
The MACHO  
\cite{kn:MACHO} collaboration 
found by now more than $\sim$ 150 
microlensing events towards the galactic bulge, most of
which are listed among the alert events, which are 
constantly updated \footnote{Current information on the MACHO
Collaboration's Alert events is maintained at the WWW site
http://darkstar.astro.washington.edu.}.
They found also
3 events by monitoring the spiral arms in the region of Gamma Scutum.
During their first season they found 45 events towards the bulge.
The MACHO team detected also in a long duration event the parallax
effect due to the motion of the Earth around the Sun \cite{kn:??}.
The MACHO first year data leads to an estimated optical depth of
$\tau \simeq 2.43^{+0.54}_{-0.45} \times 10^{-6}$, which is roughly
in agreement with the OGLE result, and which also implies the presence
of a bar in the galactic centre.
These results are very important
in order to study the structure of our Galaxy. In this respect the measurement
towards the spiral arms will give important new information.

\subsection{Towards the Andromeda galaxy}

Microlensing searches have also been conducted towards M31. In this
case, however, one has to use the so-called ``pixel-lensing'' method,
since the source stars are in general no longer resolvable. Two groups
have performed searches: the French AGAPE \cite{kn:Agape}
using the 2 meter telescope at Pic du Midi  
and the American VATT/COLUMBIA \cite{kn:VATT}, 
which used the 1.8 meter VATT-telescope
located on Mt. Graham and the 4 meter KNPO telescope.
Both teams showed that the pixel-lensing method works, however,
the small amount of observations done so far does not allow
to draw firm conclusions. The VATT/COLUMBIA
team found six candidates
which are consistent with microlending, however, additional observations
are needed to confirm this.
Pixel-lensing could also lead to the discovery of microlensing
events towards the M87 galaxy, in which case the best would be to use
the Hubble Space Telescope \cite{kn:M87}. It might also
be interesting to look towards dwarf galaxies of the local group.

\subsection{Further developments}

A new collaboration between New Zealand and Japan, called MOA, started
in june 1996 to perform observations using the 0.6 meter telescope
of the Mt. John Observatory \cite{kn:Moa}. 
The targets are the LMC and the galactic
bulge. They will in particular search for short timescale ($\sim$
1 hour) events, and would then be particularly sensitive to objects
with a mass tipical for brown dwarfs.

It has to mentioned that there are also collaborations between 
different observatories (for instance PLANET \cite{kn:PLANET}
and GMAN \cite{kn:GMAN}) 
with the aim to perform
accurate photometry on alert microlensing events. 
The GMAN collaboration was able to accurately get photometric data
on a 1995 event towards the galactic bulge. The light curve
shows clearly a deviation due to the extension of the source star
\cite{kn:gman}.
A major goal of the PLANET and GMAN collaborations 
is to find planets in binary microlensing events
\cite{kn:Mao,kn:Loeb,kn:Rhie}.
Moreover, microlensing searches are also very powerful ways to 
get large database for the study and discovery of many variable stars.

At present the only information available from a microlensing event
is the time scale, which depends on three parameters: distance,
transverse velocity and mass of the MACHO. A possible way to get more
information is to observe an event from different locations, with typically
an Astronomical Unit in separation. This could be achieved 
by putting a parallax satellite into solar orbit \cite{kn:Refsdal,kn:Gould}.

The above list of presently active collaborations and main results
shows clearly that this field is just at the beginning and that
many interesting results will come in the near future. 

\section{Basics of microlensing}

In the following we present the main features of microlensing,
in particular its probability and rate of events 
(for reviews see also \cite{kn:Pac,kn:Roulet}, whereas for double
lenses see for instance ref. \cite{kn:Dominik}).
An important issue is 
the determination from the observations of the mass of the MACHOs that
acted as gra\-vi\-tational lenses as well as the fraction of halo dark
matter they make up.
The most appropriate way to compute the average mass and other
important information is to use
the method of mass moments developed by De R\'ujula et al. \cite{kn:Derujula},
which will be briefly discussed in Section 5.6.

\subsection{Microlensing probability}

When a
MACHO of mass $M$ is sufficiently close to the line of sight
between us and a more distant
star, the light from the source suffers a gravitational
deflection. 
The deflection angle is usually so small that we do not see
two images but rather a magnification  of the original star brightness.
This magnification, at its maximum, is given by
\begin{equation}
A_{max}=\frac{u^2+2}{u(u^2+4)^{1/2}}~ . \label{eq:bb}
\end{equation}
Here $u=d/R_E$ ($d$ is the distance of the MACHO from the line of sight)
and the Einstein radius $R_E$ is defined as
\begin{equation}
R_E^2=\frac{4GMD}{c^2}x(1-x) \label{eq:cc}
\end{equation}
with $x=s/D$, and
where $D$ and $s$ are the distance between the source, respectively 
the MACHO and the observer. 

An important quantity is the optical depth $\tau_{opt}$ 
to gravitational microlensing defined as
\begin{equation}
\tau_{opt}=\int_0^1 dx \frac{4\pi G}{c^2}\rho(x) D^2 x(1-x)
\label{eq:za}
\end{equation}
with $\rho(x)$ the mass density of microlensing matter at distance
$s=xD$ from us along the line of sight. 
The quantity $\tau_{opt}$ is the probability
that a source is found within a radius $R_E$ of some MACHO and thus has a
magnification that is larger
than $A= 1.34$ ($d \leq R_E$).

We calculate $\tau_{opt}$ for a galactic mass
distribution of the form
\begin{equation}
\rho(\vec r)=\frac{\rho_0(a^2+R^2_{GC})}
{a^2+\vec r^2}~, \label{eq:zb}
\end{equation}
$\mid \vec r \mid$ being the distance from the Earth.
Here, $a$ is the core radius,
$\rho_0$ the local dark mass
density in the solar system and $R_{GC}$ the distance
between the observer and the Galactic centre.
Standard values for the
parameters are
$\rho_0=0.3~GeV/cm^3=7.9~10^{-3} M_\odot/pc^3$,
$a=5.6~kpc$ and $R_{GC}=8.5~kpc$.
With these values we get, for a spherical halo, $\tau_{opt} \simeq
5 \times 10^{-7}$
for the LMC and $\tau_{opt} \simeq 7\times 10^{-7}$ 
for the SMC \cite{kn:locarno}.

The magnification of the brightness of a star by a MACHO is a time-dependent
effect.
For a source that can be considered as
pointlike (this is the case if the projected star radius at the MACHO
distance is much less than $R_E$) 
the light curve as a function of time is obtained by inserting
\begin{equation}
u(t)=\frac{(d^2+v^2_Tt^2)^{1/2}}{R_E} \label{eq:zd}
\end{equation}
into eq.(\ref{eq:bb}), 
where $v_T$ is the transverse velocity of the MACHO, which can be inferred
from the measured rotation curve ($v_T \approx 200~ km/s$). The
achromaticity, symmetry and uniqueness of the signal are distinctive
features that allow to discriminate a microlensing event from
background events such as variable stars.

The behaviour of the magnification with time, $A(t)$, determines two
observables namely, the magnification at the peak $A(0)$ - denoted
by $A_{max}$ -
and the width of the signal $T$ (defined as 
being $T = R_E/v_T$).

\subsection{Microlensing rate towards the LMC}

The microlensing rate depends on the mass and velocity distribution of
MACHOs. 
The mass density at a distance $s=xD$ from the observer is given by
eq.(\ref{eq:zb}).
The isothermal
spherical halo model does not determine the MACHO number density as a
function of mass. A
simplifying  assumption is to let the mass distribution be independent
of the position in the galactic halo, i.e., we assume the following
factorized form for the number density per unit mass $dn/dM$,
\begin{equation}
\frac{dn}{dM}dM=\frac{dn_0}{d\mu}
\frac{a^2+R_{GC}^2}{a^2+R_{GC}^2+D^2x^2-2DR_{GC}x cos\alpha}~d\mu=
\frac{dn_0}{d\mu} H(x) d\mu~,
\label{eq:zj}
\end{equation}
with $\mu=M/M_{\odot}$ ($\alpha$ is the angle of the 
line of sight with 
the direction of the galactic centre, which is $82^0$ for
the LMC), $n_0$ not depending on $x$ 
and is subject to the normalization
$\int d\mu \frac{dn_0}{d\mu}M=\rho_0$.
Nothing a priori is known on the distribution $d n_0/dM$.

A different situation arises for the velocity
distribution in the isothermal
spherical halo model, its
projection in the plane perpendicular to the line of sight
leads to the following
distribution in the transverse velocity $v_T$
\begin{equation}
f(v_T)=\frac{2}{v_H^2}v_T e^{-v^2_T/v_H^2}.\label{eq:zr}
\end{equation}
($v_H \approx 210~km/s$ is the observed velocity dispersion in the halo).

In order to find the rate at which a single star
is microlensed with magnification
$A \geq A_{min}$, we consider MACHOs
with masses between $M$ and $M+\delta M$, located at a distance from
the observer between $s$ and $s+\delta s$ and with transverse velocity
between $v_T$ and $v_T+\delta v_T$. The collision time can be
calculated using the well-known fact that the inverse of the collision
time is the product of the MACHO number density, the microlensing
cross-section and the velocity. 
The rate $d\Gamma$, taken also as a differential with respect 
to the variable $u$, at which a single star is microlensed
in the interval $d\mu du dv_T dx$ is given by
\cite{kn:Derujula,kn:Griest1}
\begin{equation}
d\Gamma=2v_T f(v_T)D r_E [\mu x(1-x)]^{1/2} H(x)
\frac{d n_0}{d\mu}d\mu du dv_T dx,\label{eq:zt}
\end{equation}
with
\begin{equation}
r_E^2=\frac{4GM_{\odot}D}{c^2} \sim
(3.2\times 10^9 km)^2 .\label{eq:zs}
\end{equation}

One has to integrate
the differential number of microlensing events, 
$dN_{ev}=N_{\star} t_{obs} d\Gamma$,
over an appropriate range for $\mu$, $x$,
$u$ and $v_T$, 
in order to obtain the total number of microlensing events which can
be compared with an experiment
monitoring $N_{\star}$ stars during an
observation time $t_{obs}$ and which is able to detect
a magnification such that $A_{max} \geq A_{TH}$.
The limits of the $u$ integration are determined by
the experimental threshold in magnitude shift, $\Delta m_{TH}$:
we have $0 \leq u \leq u_{TH}$.

The range of integration for $\mu$ is where the mass
distribution $dn_0/d\mu$ is not vanishing
and that for $x$ is
$0\leq x \leq D_h/D$ where $D_h$ is the extent of the galactic halo along
the line of sight (in the case of the LMC,
the star is inside the galactic halo and thus $D_h/D=1$.)
The galactic velocity distribution is cut at the escape velocity
$v_e \approx 640~km/s$ and therefore
$v_T$ ranges over $0\leq v_T \leq v_e$.
In order to simplify the integration we integrate $v_T$
over all the positive axis, due to the exponential factor in $f(v_T)$ the
so committed error is negligible.

However, the integration range of $d\mu du dv_T dx$
does not span all the interval we have just described.
Indeed, each experiment has time
thresholds $T_{min}$ and $T_{max}$ and only detects events with:
$T_{min}\leq T \leq T_{max}$,
and thus the integration range has to be such that $T$ lies in this
interval.
The total number of micro-lensing events is then given by
\begin{equation}
N_{ev}=\int dN_{ev}~\epsilon(T)~
,\label{eq:th}
\end{equation}
where the integration is over the full range of
$d\mu du dv_T dx$. $\epsilon(T)$ is determined 
experimentally \cite{kn:Pratt,kn:MACHO}
or if not known can, for instance, be taken as follows
$\epsilon(T) =\Theta (T-T_{min})\Theta (T_{max}-T)$.
$T$ is related in a complicated way
to the integration variables,
because of this, no direct
analytical integration in eq.(\ref{eq:th}) can be performed.

To evaluate eq.(\ref{eq:th}) we define
an efficiency function $\epsilon_0(\mu)$
\begin{equation}
\epsilon_0(\mu) \equiv \frac{\int d N^{\star}_{ev}(\bar\mu)~ 
\epsilon(T)}
{\int d N^{\star}_{ev}(\bar\mu)}~,
\end{equation}
which measures the fraction of the total number of microlensing events
that meet the condition on $T$ at a
fixed MACHO mass $M=\bar\mu M_{\odot}$.
We now can write the total number of events in
eq.(\ref{eq:th}) as
\begin{equation}
N_{ev}=\int dN_{ev}~\epsilon_0(\mu)~.\label{eq:tl}
\end{equation}
Due to the fact that
$\epsilon_0$ is a function of $\mu$ alone, the integration in
$d\mu du dv_T dx$ factorizes into four integrals with independent
integration limits. 

The average lensing duration can be defined as follows
\begin{equation}
< T > = \frac{1}{\Gamma}~\int d\Gamma~T(x,\mu,v_T)~,
\end{equation}
where $T(x,\mu,v_T) = R_E(x,\mu)/v_T$. One easily finds that $< T >$
satisfies the following relation
\begin{equation}
< T > = \frac{2 \tau}{\pi \Gamma} ~u_{TH}~.
\end{equation}

In order to quantify the expected number of events it is convenient
to take as an example a delta function distribution for the mass.
The rate of microlensing
events with
$A \geq A_{min}$ (or $u \leq u_{max}$), is then
\begin{equation}
\Gamma(A_{min})=u_{max} \tilde\Gamma = u_{max} 
D r_E \sqrt{\pi}~v_H \frac{\rho_0}{M_{\odot}}\frac{1}{\sqrt{\bar \mu}}
\int^1_0 dx[x(1-x)]^{1/2} H(x)~.\label{eq:ta}
\end{equation}

Inserting the numerical values for the LMC
(D=50~kpc and $\alpha=82^0$) we get
\begin{equation}
\tilde\Gamma=4
\times 10^{-13}~\left( \frac{v_H}{210~km/s}\right)
  \left( \frac{\rho_0}{0.3~GeV/cm^3}\right)
\frac{1}{\sqrt{M/M_{\odot}}}\ ~{\rm s^{-1}}.
\label{eq:tb}
\end{equation}
For an experiment monitoring $N_{\star}$ stars during an
observation time $t_{obs}$ the total number of events with a
magnification $A \geq A_{min}$ is:
$N_{ev}(A_{min})=N_{\star} t_{obs} \Gamma(A_{min})$.
In the following Table 1 we show some values of $N_{ev}$ for the LMC,
taking
$t_{obs}=1$ year, $N_{\star}=10^6$ stars and 
$A_{min} = 1.34$ (or $\Delta m_{min} = 0.32$).
\vskip 0.3 cm 
Table 1
\vskip 0.2 cm
\begin{center}
\begin{tabular}{|c|c|c|c|}\hline
MACHO mass in $M_{\odot}$ & Mean $R_E$ in km & Mean microlensing time &
$N_{ev}$ \\
\hline
$10^{-1}$ & $0.3\times 10^9$ & 1 month & 4.5  \\
$10^{-2}$ & $10^8$ & 9 days & 15 \\
$10^{-4}$ & $10^7$ & 1 day & 165 \\
$10^{-6}$ & $10^6$ & 2 hours & 1662 \\
\hline
\end{tabular}
\end{center}

\vskip 0.2 cm
As mentioned in Sect. 4.1 the MACHO team found (till September 1997)
altogether 14 events (one of which being a binary lens) towards the
LMC and one event towards the SMC. 
The EROS team found 2 events using photographic plate techniques.

\subsection{Microlensing rate towards M31}

Gravitational microlensing is also useful for detecting MACHOs in
the halo of nearby galaxies \cite{kn:Crotts,kn:Baillon} 
such as M31 or M33, for which the experiments AGAPE and VATT/Columbia
are under way.
In fact, it turns out
that the massive dark halo of M31 has an optical depth to microlensing
which is of about the same order of magnitude as that of our own
galaxy \cite{kn:Crotts,kn:Baillon,kn:Jetzer} $\sim 10^{-6}$. 
Moreover, an experiment monitoring stars in
M31 is sensitive to both MACHOs in our halo and in the one of M31.
One can also compute the microlensing rate \cite{kn:Jetzer}
for MACHOs in the halo of M31, for which we get 
\begin{equation}
\tilde\Gamma=1.8 \times 10^{-12} \left(\frac{v_H}{210~km/s} \right)
\left(\frac{\rho(0)}{1~Gev/cm^3} \right)
\frac{1}{\sqrt{M/M_{\odot}}}~{\rm s^{-1}}~. \label{eq:tc1}
\end{equation}
($\rho(0)$ is the central density of dark matter.)
The average lensing time is given by
\begin{equation}
< T > \sim (125~ days)~ \sqrt{M/M_{\odot}}~. \label{eq:tc}
\end{equation}

In the following Table 2 we show some values of $N^a_{ev}$ due to MACHOs in the
halo of M31 with $t_{obs}= 1$ year and $N_{\star}=10^6$ stars. In the 
last column we give the corresponding number of events, $N_{ev}$,
due to MACHOs in our    
own halo. The mean microlensing time is about the same for both types of 
events.
\newpage
Table 2
\vskip 0.2 cm
\begin{center}
\begin{tabular}{|c|c|c|c|c|}\hline
MACHO mass in $M_{\odot}$ & Mean $R_E$ in km & Mean microlensing time &
$N^a_{ev}$ & $N_{ev}$\\
\hline
$10^{-1}$ & $7\times 10^8$ & 38 days & 6 & 3 \\
$10^{-2}$ & $2\times 10^8$ & 12 days & 21 & 12 \\
$10^{-4}$ & $2\times 10^7$ & 30 hours & 210 & 129 \\
$10^{-6}$ & $2\times 10^6$ & 3 hours & 2100 & 1290 \\      
\hline
\end{tabular}
\end{center}

\vskip 0.2 cm

$N_{ev}^a$ is almost by
a factor of two bigger than
$N_{ev}$ (see also \cite{kn:Jetzer}). 
Of course these numbers should be taken as an estimate, since they
depend on the details of the  model
one adopts for the distribution of  
the dark matter in the halo.
To distinguish between events due to a MACHO in our halo or in the one of
M31 might be rather difficult.

\subsection{Microlensing rate towards the galactic bulge}

We compute the microlensing rate $\Gamma$
for an experiment monitoring
stars in the galactic bulge in Baade's window of galactic
coordinates (longitude and latitude): $l = 1^{\circ}$, $b = -3.^{\circ}9$.
$D = 8.5$ kpc is the distance
to stars in the galactic bulge, $r_E=1.25 \times 10^9$ km
and we use the definition $T=R_E/v_T$.
 
The number density of disk stars per unit mass is given by
\cite{kn:Bahcall}
\begin{equation}
\frac{dn}{dM}=\frac{dn_0}{dM}~
\exp\left(-\frac{D x \mid \sin~b \mid}{300~{\rm pc}}
+\frac{D x~\cos~b}{3.5~{\rm kpc}}\right)
=\frac{dn_0}{dM}~H_d(x)~, \label{eqno:3a}
\end{equation}
where the galactic longitude $l=0^{\circ}$ has been adopted and
$\frac{dn_0}{dM}=\frac{\rho_d(M)}{M}$ with $\rho_d(M)=0.05 M_{\odot}~
{\rm pc}^{-3}$; $H_d$ can be written as follows
\begin{equation}
H_d(x)=\exp\left(\frac{x D~ (1-11.7
\mid \sin~b \mid)}{3.5~{\rm kpc}}\right)~,
\label{eqno:3b}
\end{equation}
using also the fact that $\cos~b \approx 1$ for the galactic bulge.
 
For the contribution $H(x)$ of the halo dark matter in form of MACHOs
located in the disk one uses the distribution
for the number density per unit mass $dn/dM$ as given in eq. (\ref{eq:zj})
with $\alpha \simeq~ 4^{\circ}$ 
(the angle between the line of sight and the direction of the galactic
centre).

With the above distribution for the lenses in the disk it follows that
the corresponding optical depth is $\tau \simeq 7 \times 10^{-7}$
(and $\tau \simeq 1.2 \times 10^{-7}$ for the halo contribution).
 
In computing $\Gamma$ one must also take
into account the fact that both the source and
the observer are in motion \cite{kn:Griest1}.
Relevant are only the velocities transverse
to the line of sight.
The transverse
velocity of the microlensing tube at position $xD$ is:
$\vec v_t(x)=(1-x)\vec v_{\odot \perp}+x \vec v_{s \perp}$,
and its magnitude is 
\begin{equation}
v_t(x)=\sqrt{(1-x)^2 \mid \vec v_{\odot \perp} \mid^2+x^2\mid \vec
v_{s \perp}\mid^2+2x(1-x)\mid \vec v_{\odot \perp}\mid \mid
\vec v_{s \perp}
\mid \cos~\theta }~, \label{eqno:5a}
\end{equation}
where $\vec v_{s \perp}$ and $\vec v_{\odot \perp}$ are
the source and the
solar velocities transverse to the line of sight and $\theta$ the
angle between them.
 
For the velocity distribution of the MACHOs or the faint disk stars we
consider an isothermal spherical model, which in the rest
frame of the galaxy is given by
\begin{equation}
f(\vec v) d^3v=\frac{1}{\tilde v^3_H \pi^{3/2}}~
e^{-\vec v^2/\tilde v^2_H}~d^3v~ . \label{eqno:4}
\end{equation}
Since only the transverse velocities are of relevance cylindrical
coordinates can be used
and the integration made over the velocity component parallel to the
line of sight. Moreover, due to the velocities of the observer and the
source, the value of the transverse velocity gets shifted
by $\vec v_t(x)$.
The distribution for the transverse velocity is thus
\begin{equation}
\tilde f(v_T) dv_T=\frac{1}{\pi v^2_H}~
e^{-(\vec v_T-\vec v_t)^2/v^2_H}~v_T~dv_T~,
\label{eqno:4b}
\end{equation}
where $v_H \approx 30~{\rm km}~{\rm s}^{-1}$ 
is the velocity dispersion \cite{kn:Paczynski1991}.

The random velocity of the source stars in the bulge are
again described by an
isothermal spherical distribution, whose transverse
velocity distribution is
\begin{equation}
g(v_{s \perp})dv_{s \perp}=\frac{1}
{\pi v^2_D}~e^{-v^2_{s \perp}/v^2_D}~v_{s \perp}~
dv_{s \perp}~,  \label{eqno:4c}
\end{equation}
where $v_D=156~{\rm km}~{\rm s}^{-1}$
is the velocity dispersion \cite{kn:Mihalas,kn:Paczynski1991,kn:Griest2}.
 
Taking all the above facts into account,
$\Gamma$ turns out to be \cite{kn:Jetzer1}
\begin{eqnarray}
\Gamma &=& 4~r_E D~v_H~u_{TH} \left( \int_0^{\infty}
\sqrt{\mu}~\frac{dn_0}{d\mu}~d\mu
\right) \int_0^{2\pi} d\theta ~\int_0^{\infty} dv_{s \perp}~
g(v_{s \perp}) \nonumber \\ & &
~\int_0^1 \sqrt{x(1-x)}~H_i(x)~e^{-\eta^2}
\int_0^{\infty} dy~y^2~I_0(2y\eta)~e^{-y^2}~, \label{eqno:4d}
\end{eqnarray}
where $y=\frac{v_T}{v_H}$, $\eta(x,\theta,v_{s \perp},v_{\odot \perp})=
\frac{v_t}{v_H}$, and $I_0$ is the modified Bessel function of order 0.
In the limit of stationary observer and source star ($v_{\odot \perp}=
v_{s \perp}=0)$,
$\eta=0$ and $I_0=1$; $H_i$ means either $H_d$ or
$H$; $u_{TH}$ is related to the minimal experimentally detectable
magnification $A_{TH}=A[u=u_{TH}]$; $v_{\odot \perp}$ is
$\approx 220~{\rm km}~{\rm s}^{-1}$ (more precisely it should
be multiplied by $\cos~ l$,
where $l$ is the galactic longitude
but since $l = 1^{\circ}$, $\cos~ l \approx 1$).
In computing $\Gamma$ one should also take into account the limited
measurable range for the event duration $T$, which translates into a
modification of the integration limits. A fact that can be
described by introducing an efficiency function \cite{kn:Derujula}
$\epsilon_0(\mu)$. 
 
Assuming a delta-function-type distribution for the masses
\begin{equation}
\frac{dn_0}{d\mu}=\frac{\rho}{M_{\odot}}~\frac{\delta(\mu-\bar\mu)}
{\mu}~, \label{eqno:4da}
\end{equation}
eq. (\ref{eqno:4d}) can be integrated.
With $N_{\star}=10^{6}$ stars and $t_{obs}=1$ year one gets
\begin{equation}
N_{ev}= \frac{2.08}{\sqrt{\bar\mu}}~\left(\frac{\rho_d}{5 \times
10^{-2} M_{\odot}~
{\rm pc}^{-3}} \right)~u_{TH}~, \label{eqno:4e}
\end {equation}
for $H_i=H_d$, and
\begin{equation}
N_{ev}=\frac{0.61}{\sqrt{\bar\mu}}~
\left(\frac{\rho_0}{8 \times 10^{-3} M_{\odot}
~{\rm pc}^{-3}} \right)~u_{TH}~, \label{eqno:4f}
\end{equation}
for $H_i=H$. The numerical factor in
eq. (\ref{eqno:4e}) for
$N_{ev}$ as a function of $b$, the galactic latitude,
varies between 6.8 for $b=0^{\circ}$ and 1.9 for $b=5^{\circ}$, 
whereas the factor
for $N_{ev}$ of eq. (\ref{eqno:4f}) remains practically unchanged.

As mentioned in Section 4.2 the results clearly show that there is a
bar in the galactic centre and that one has to consider also the 
bar-bar (or bulge-bulge) contribution in order to explain the observations. 

\subsection{Most probable mass for a single event}

The probability $P$ that a microlensing
event of duration $T$
and maximum amplification $A_{max}$ be produced by a MACHO
of mass $\mu$ (in units of $M_{\odot}$) is given by
\cite{kn:Jetzer2} 
\begin{equation}
P(\mu,T) \propto \frac{\mu^2}{ T^4} \int_0^1 dx (x(1-x))^2 H(x)
exp\left( -\frac{r_E^2 \mu x(1-x)}{v^2_H  T^2} \right) ~, \label{eqno:81}
\end{equation}
which does not dependent on $A_{max}$ 
and $P(\mu, T)=P(\mu/ T^2)$. The measured values for
$ T$ towards the LMC are listed in Table 3, where 
$\mu_{MP}$ is
the most probable value.
The normalization
is arbitrarily
chosen such that the maximum of $P(\mu_{MP}, T)=1$. 
We find that the maximum corresponds to 
$\mu r_E^2/v^2_H T^2=13.0$ \cite{kn:Jetzer2,kn:Jetzer1}. 
The 50\% confidence interval
embraces for the mass $\mu$ approximately
the range $1/3\mu_{MP}$ up to $3 \mu_{MP}$.
Similarly one can compute $P(\mu, T)$ also for the bulge events
\cite{kn:Jetzer1}.

\vskip 0.3cm 
Table 3: Values of $\mu_{MP}$ (in $M_{\odot}$)
for eight microlensing events detected in the LMC ($A_{i}$
= American-Australian
collaboration events ($i$ = 1,..,6);
$F_1$ and $F_2$
French collaboration events).
For the LMC: $v_H = 210~{\rm km}~{\rm s}^{-1}$ and
$r_E = 3.17 \times 10^9~{\rm km}$.
\vskip 0.2cm
\begin{center}
\begin{tabular}{|c|c|c|c|c|c|c|c|c|}\hline
  & $A_{1}$ & $A_{2}$ & $A_3$ & $A_4$ & $A_5$ & $A_6$ & $F_1$ & $F_2$  \\
\hline
$ T$ (days) & 17.3 & 23 & 31 & 41 & 43.5 & 57.5 & 27 & 30 \\
\hline
$\tau (\equiv \frac{v_H}{r_E} T)$ & 0.099 &
0.132 & 0.177 & 0.235 & 0.249 & 0.329 & 0.155 & 0.172 \\
\hline
$\mu_{MP}$ & 0.13 & 0.23 & 0.41 & 0.72 & 0.81 & 1.41 & 0.31 & 0.38 \\
\hline
\end{tabular}
\end{center} 

\subsection{Mass moment method}

A more systematic way to extract information on the masses is to use the
method of mass moments \cite{kn:Derujula}. 
The mass moments $<\mu^m>$ are defined as
\begin{equation}
<\mu^m>=\int d\mu~ \epsilon_n(\mu)~ 
\frac{dn_0}{d\mu}\mu^m~. \label{eqno:10}
\end{equation}
$<\mu^m>$ is related to $<\tau^n>=\sum_{events} \tau^n$,
with $\tau \equiv (v_H/r_E) T$, as constructed
from the observations and which can also be computed as follows
\begin{equation}
<\tau^n>=\int dN_{ev}~ \epsilon_n(\mu)~
\tau^n=V u_{TH} ~\gamma(m) <\mu^m>~,
\label{eqno:111}
\end{equation}
with $m \equiv (n+1)/2$.
For targets in the LMC $\gamma(m) = \Gamma(2-m) \widehat H(m)$ and
\begin{equation}
V \equiv 2 N_{\star} t_{obs}~ D~ r_E~ v_H=2.4 \times 10^3~ pc^3~ 
\frac{N_{\star} ~t_{obs}}{10^6~ {\rm star-years} }~, \label{eqno:121}
\end{equation}
\begin{equation}
\Gamma(2-m) \equiv \int_0^{\infty} \left(\frac{v_T}{v_H}\right)^{1-n}
f(v_T) dv_T~,
\label{eqno:131}
\end{equation}
\begin{equation}
\widehat H(m) \equiv \int_0^1 (x(1-x))^m H(x) dx~.  \label{eqno:14}
\end{equation}
The efficiency $\epsilon_n(\mu)$ is determined as follows 
(see \cite{kn:Derujula})
\begin{equation}
\epsilon_n(\mu) \equiv \frac{\int d N^{\star}_{ev}(\bar\mu)~ 
\epsilon(T)~ \tau^n}
{\int d N^{\star}_{ev}(\bar\mu)~ \tau^n}~, \label{eqno:15}
\end{equation}
where $d N^{\star}_{ev}(\bar\mu)$ is defined as $d N_{ev}$ 
in eq.(\ref{eq:th}) with
the MACHO mass distribution concentrated at a fixed mass
$\bar\mu$: $dn_0/d\mu=n_0~ \delta(\mu-\bar\mu)/\mu$. 
$\epsilon(T)$ is the experimental detection efficiency.
For a more detailed discussion on the efficiency see ref. \cite{kn:Masso}.

A mass moment $< \mu^m >$ is thus related to 
$< \tau^n >$ as given from the measured values 
of $T$ in a microlensing experiment by
\begin{equation}
< \mu^m > = \frac{< \tau^n >}{V u_{TH} \gamma(m)}~.
\label{eqno:16}
\end{equation}

The mean local density of MACHOs (number per cubic parsec)
is $<\mu^0>$. The average local mass density in MACHOs is
$<\mu^1>$ solar masses per cubic parsec. 
In the following we consider only 6 (see Table 3)
out of the 8 events observed by the MACHO group,
in fact the two events we neglect are 
a binary lensing event and an event which is rated as marginal.
The  mean mass, which we get from
the six events detected by the MACHO team, is \cite{kn:je} 
\begin{equation}
\frac{<\mu^1>}{<\mu^0>}=0.27~M_{\odot}~.
\label{eqno:aa}
\end{equation}
(To obtain this result we used the values of $\tau$
as reported in Table 3, whereas $\Gamma(1)\widehat H(1)=0.0362$ and
$\Gamma(2)\widehat H(0)=0.280$ as
plotted in Fig. 6 of ref. \cite{kn:Derujula}).
When taking for the duration $T$ the values corrected for ``blending'',
we get as average mass 0.34 $M_{\odot}$.
If we include also the two EROS events we get a value
of 0.26 $M_{\odot}$ for the mean mass (without taking into account
blending effects).
The resulting mass depends on the parameters
used to describe the standard halo model. In order to check this
dependence we varied the parameters within
their allowed range and found
that the average mass changes at most by $\pm$ 30\%, which shows
that the result is rather robust. 
Although the value for the average mass we find with the mass moment
method is marginally consistent with the result of the MACHO team,
it definitely favours a lower average MACHO mass.

One can
also consider other models with more general
luminous and dark matter distributions, e.g. ones with a flattened halo
or with anisotropy in velocity space \cite{kn:Ingrosso},
in which case the resulting
value for the average mass would decrease significantly.

Another important quantity to be determined is the fraction $f$ of the local
dark mass density (the latter one given by $\rho_0$) detected
in the form of MACHOs, which is given by
$f \equiv {M_{\odot}}/{\rho_0} \sim 126~{\rm pc}^3$ $<\mu^1>$.
Using the values given by the MACHO collaboration
for their two years data \cite{kn:Pratt} (in particular
$u_{TH}=0.661$ corresponding to $A > 1.75$ and
an effective exposure $N_{\star} t_{obs}$
of $\sim 5 \times 10^6$ star-years for 
the observed range of the event duration $T$ between $\sim$ 20 - 50 days)
we find $f \sim 0.54$, which compares quite well
with the corresponding value ($f \sim 0.45$ based on the six events
we consider) calculated
by the MACHO group in a different way. The value for $f$ is obtained 
again by assuming a standard spherical halo model.

Similarly, one can also get information from the events
detected so far towards the galactic bulge.
The mean MACHO mass, which one gets when considering
the first eleven events detected by OGLE in the galactic bulge,
is $\sim 0.29 M_{\odot}$ \cite{kn:Jetzer1}.
From the 40 events discovered 
during the first year of operation
by the MACHO team \cite{kn:MACHO} (we considered
only the events used by the MACHO team to infer the optical
depth without the double lens event)
we get an average value
of 0.16$M_{\odot}$. Both values are obtained under the assumption that
the lenses are located in the disk. For a more detailed analysis see
ref \cite{kn:Diploma}.
The lower value inferred from the MACHO data is due to the fact
that the efficiency for the short duration events ($\sim$ some days)
is substantially higher for the MACHO experiment than for the
OGLE one. 
These values for the average mass
suggest that the lens are faint stars. 

Once several moments $< \mu^m >$ are known one can
get information on the mass distribution $dn_0/d\mu$. 
Since at present only few events towards the LMC are at disposal the 
different moments (especially the higher ones) can 
be determined only approximately.
Nevertheless, the results obtained so far
are already of interest and it is clear that in a few years,
due also to the new experiments under way (such as EROS II, OGLE II
and MOA in addition to MACHO),
it will be possible to draw more firm conclusions.

\section{Formation of dark clusters}

A major problem concerns the formation of MACHOs, as well
as the nature of the remaining amount of dark matter in the galactic halo.
We feel it hard to conceive a formation mechanism which transforms with 100\%
efficiency hydrogen and helium gas into MACHOs. Therefore, we expect
that also cold clouds (mainly of $H_2$) should be present in the 
galactic halo. Recently, we have proposed
a scenario \cite{de,de1,de2,de3,de4}
in which dark clusters of MACHOs and cold molecular 
coulds naturally form in the halo at galactocentric distances
larger than 10--20 kpc, with the relative abundance possibly
depending on the distance. 
Similar scenario have also been considered in refs. 
\cite{kn:Pfenniger,kn:Silk1}.

The evolution of the primordial proto globular cluster clouds
(which make up the proto-galaxy)
is expected to be very different 
in the inner and outer parts of the Galaxy, depending on the decreasing 
ultraviolet flux (UV) from the centre
as the galactocentric distance $R$ increases.
In fact, in the outer halo no substantial $H_2$ depletion should 
take place, owing to the distance suppression of the UV flux.
Therefore, the clouds cool and fragment - the process stops when the 
fragment mass becomes $\sim 10^{-2} - 10^{-1}~M_{\odot}$.
In this way dark
clusters should form, which contain brown dwarfs  
and also cold $H_2$ self-gravitating cloud, 
along with some residual diffuse gas (the 
amount of diffuse gas inside a dark cluster has to be low, for otherwise it 
would have been observed in the radio band).

We have also considered several observational tests for our model 
\cite{de,di}.
In particular, a signature for the presence of 
molecular clouds in the galactic halo should be a $\gamma$-ray flux 
produced in the scattering of high-energy cosmic-ray protons on $H_2$.
As a matter of fact, an essential
information is the knowledge of the cosmic ray flux in the halo. Unfortunately,
this quantity is unknown and the only available 
information comes from theoretical considerations.
Nevertheless, we can make an estimate of the expected $\gamma$-ray flux
and the best chance to detect it is provided
by observations at high galactic latitude.
Accordingly, we find a $\gamma$-ray flux (for $E_{\gamma}>100$ MeV)
$\Phi_{\gamma}(90^0) \simeq ~\tilde f~(0.4 - 1.8) \times 10^{-5}$ 
photons cm$^{-2}$
s$^{-1}$  sr$^{-1}$ ($\tilde f$ stands for 
the fraction of halo dark matter in the form of gas),
if the cosmic rays are confined in the galactic halo, otherwise, if they
are confined in the local galaxy group \cite{kn:ber} 
$\Phi_{\gamma}(90^0) \simeq ~\tilde f~(0.6 - 3) \times 10^{-7}$ 
photons cm$^{-2}$
s$^{-1}$  sr$^{-1}$. 
These values should be compared with the measured
flux by the SAS-II satellite for the diffuse background of
$(0.7-2.3)\times 10^{-5}$ photons cm$^{-2}$ s$^{-1}$  sr$^{-1}$
or the corresponding flux found by EGRET of 
$\sim 1.1 \times 10^{-5}$ photons cm$^{-2}$ s$^{-1}$  sr$^{-1}$. Thus, 
there is at present no contradiction with observations.
Furthermore, an improvement of sensitivity for the next generation of 
$\gamma$-ray detectors will allow to clarify the origin
of this flux   
or yield more stringent limits on $\tilde f$.\\

\section{Conclusions}

The mistery of the dark matter is still unsolved, however, thanks
to the ongoing microlensing and pixel-lensing
experiments there is hope that
progress on its nature in the galactic halo
can be achieved within the next few years. 
Substantial progress will also be done in the study of the
structure of our Galaxy and this especially once data from the
observations towards the spiral arms will be available.
Microlensing is also very promising for the discovery of planets.
Although being a rather young observational technique
microlensing has
already allowed to make substantial progress and
the prospects for further contribution
to solve important astrophysical problems look very bright. 

It has also to be mentioned that
it is well plausible that only a fraction of the halo
dark matter is in form of MACHOs, either brown dwarfs or white
dwarfs, in which case there is the problem of explaining the nature
of the remaining dark matter and the formation of the MACHOs.
Before invoking the need for new particles
as galactic dark matter candidates
for the remaining fraction, one should seriously consider the
possibility that it is in the form of cold molecular clouds.
A scenario this, for which several observational tests
have been proposed, thanks to which it should be feasible
in the near future 
to either detect or to put stringent limits on these clouds.


\begin{thebibliography}{99}
\bibitem{kn:Faber}
S.M. Faber and J.S. Gallagher, Ann. Rev. Astron. Astrophys. {\bf 17}
(1979) 135
\bibitem{kn:Trimble}
V. Trimble, Ann. Rev. Astron. Astrophys. {\bf 25} (1987) 425
\bibitem{kn:Oort}
J.H. Oort, Bull. Astron. Inst. Netherlands {\bf 6} (1932) 249
\bibitem{kn:Zwicky}
F. Zwicky, Helv. Phys. Acta {\bf 6} (1933) 110
\bibitem{kn:Rubin}
V.C. Rubin and W.K. Ford, Astrophys. J. {\bf 159} (1970) 379
\bibitem{kn:Persic}
M. Persic, P. Salucci and F. Stel, Mont. Not. R. Astr. Soc.
{\bf 281} (1996) 27
\bibitem{kn:Zaritsky}
D. Zaritsky et al., Astrophys. J. {\bf 345} (1989) 759
\bibitem{kn:Lin}
D.N. Lin, B.F. Jones and A.R. Klemola, Astrophys. J. {\bf 439} (1995)
652
\bibitem{kn:Kochanek}
C.S. Kochanek, Astrophys. J. {\bf 457} (1996) 228
\bibitem{kn:jungman}
G. Jungman, M. Kamionkowski and K. Griest, Phys. Rept. {\bf 267} (1996)
195
\bibitem{kn:moore} B. Moore, Nature {\bf 370} (1994) 629
\bibitem{kn:navarro} J.F. Navarro, C.S. Frenk and S.D. White,
Astrophys. J. {\bf 462} (1996) 563
\bibitem{kn:Silk} A. Burkert and J. Silk, astro-ph 9707343
\bibitem{kn:copi}
C.J. Copi, D.N. Schramm and M.S. Turner, Science {\bf 267} (1995) 192
\bibitem{kn:PDG}
Particle Data Group, Phys. Rev. {\bf D54} (1996) 109-111
\bibitem{kn:corx}
F. De Paolis, G. Ingrosso, Ph. Jetzer and M. Roncadelli,
astro-ph 9709052 to appear in Astron. and Astrophys.
\bibitem{kn:Carr}
B. Carr, Annu. Rev. Astron. Astrophys. {\bf 32} (1994) 531
\bibitem{kn:Derujula1} A. De R\'ujula, Ph. Jetzer and E. Mass\'o,
Astron. and Astrophys. {\bf 254} (1992) 99
\bibitem{kn:JBahcall}
J. Bahcall, C. Flynn, A. Gould and S. Kirhakos, Astrophys. J.
{\bf 435} (1994) L51
\bibitem{kn:Kerins}
E. J. Kerins, astro-ph 9610070
\bibitem{kn:Kerins1}
E. J. Kerins, astro-ph 9704179
\bibitem{kn:Tamanaha}
C.M. Tamanaha, J. Silk, M.A. Wood and D.E. Winget, Astrophys. J.
{\bf 358} (1990) 164
\bibitem{kn:Kawaler}
S.D. Kawaler, Astrophys. J. {\bf 467} (1996) L61 
\bibitem{kn:Paczynski} B. Paczy\'nski, Astrophys. J. {\bf 304} (1986) 1
\bibitem{kn:Paczynski1991} B. Paczy\'nski, Astrophys. J. {\bf 371}
(1991) L63
\bibitem{kn:Griest2}
Griest, K. et al., 1991, Ap. J. {\bf 372}, L79
\bibitem{kn:Aubourg} E. Aubourg et al., Nature {\bf 365} (1993) 623
\bibitem{kn:Alcock} C. Alcock et al., Nature {\bf 365} (1993) 621;
Astrophys. J. {\bf 445}, (1995) 133
\bibitem{kn:Pratt} C. Alcock et al., astro-ph 9606165
\bibitem{kn:EROS} C. Renault et al., Astron. and Astrophys. {\bf 324} (1997)
L69
\bibitem{kn:Alcock2} C. Alcock et al., astro-ph 9708190 
\bibitem{kn:Udalski} A. Udalski et al., Acta Astron. {\bf 43} (1993) 289
\bibitem{kn:udal} A. Udalski et al., Acta Astron. {\bf 44} (1994) 165 
\bibitem{kn:Kiraga} M. Kiraga and B. Paczy\'nski, Astrophys. J. {\bf 430}
(1994) 101
\bibitem{kn:Alard} C. Alard, in Proceedings of the 12th IAP Astrophysics 
Colloquium, Editions Fronti\`eres (1997) 37
\bibitem{kn:MACHO} C. Alcock et al., Astrophys. J. {\bf 479} (1977) 119
\bibitem{kn:??} C. Alcock et al., Astrophys. J. {\bf 454} (1995) L125
\bibitem{kn:Agape} 
R. Ansari et al., Astron. and Astrophys. {\bf 324} (1997) 843.
\bibitem{kn:VATT} A.P.S. Crotts and A.B. Tomaney, Astrophys. J. {\bf 473}
(1996) L87
\bibitem{kn:M87} A. Gould, Astrophys. J. {\bf 455} (1995) 44
\bibitem{kn:Moa} F. Abe et al., in Proceedings of the 12th IAP Astrophysics 
Colloquium, Editions Fronti\`eres (1997) 75 
\bibitem{kn:PLANET} 
M. Albrow et al., in Proceedings of the IAU Symposium 173 -
Astrophysical Applications of Gravitational Lensing (Melbourne,
Australia), C.S. Kochanek and J.N. Hewitt editors,
Kluwer, Dordrecht (1996), page 227
\bibitem{kn:GMAN} Pratt et al., astro-ph 9508039
\bibitem{kn:gman} C. Alcock et al., astro-ph 9702199
\bibitem{kn:Mao} S. Mao and B. Paczy\'nski, Astrophys. J. {\bf 374} (1991)
L37
\bibitem{kn:Loeb} A. Gould and A. Loeb, Astrophys. J. {\bf 396} (1992) 104
\bibitem{kn:Rhie} D. Bennett and S.H. Rhie, astro-ph 9603158
\bibitem{kn:Refsdal} S. Refsdal, Mont. Not. R. Astr. Soc. {\bf 134} (1966)
315
\bibitem{kn:Gould} A. Gould, Astrophys. J. {\bf 421} (1994) L75x
\bibitem{kn:Pac}
B. Pacz\'ynski, Annu. Rev. Astron. Astrophys. {\bf 34} (1996) 419 
\bibitem{kn:Roulet}
E. Roulet and S. Mollerach, Phys. Rept. {\bf 279} (1997) 67
\bibitem{kn:Dominik} M. Dominik, Thesis University of Dortmund (1996)
\bibitem{kn:Derujula} A. De R\'ujula, Ph. Jetzer and E. Mass\'o,
Mont. Not. R. Astr. Soc. {\bf 250} (1991) 348
\bibitem{kn:locarno}
Ph. Jetzer, Atti del Colloquio di Matematica (CERFIM) {\bf 7}
(1991) 259
\bibitem{kn:Griest1} K. Griest, Astrophys. J. {\bf 366} (1991) 412 
\bibitem{kn:Crotts} A.P. Crotts, Astrophys. J. {\bf 399} (1992) L43
\bibitem{kn:Baillon} P. Baillon, A. Bouquet, Y. Giraud-H\'eraud and
J. Kaplan, Astron. and Astrophys. {\bf 277} (1933) 1;
\bibitem{kn:Jetzer} Ph. Jetzer, Astron. and Astrophys.
{\bf 286} (1994) 426
\bibitem{kn:Bahcall}
Bahcall, J.N., \& Soneira, R.M., 1980, Ap. JS. {\bf 44}, 73.
\bibitem{kn:Mihalas}
Mihalas, D., \& Binney, J., 1981, Galactic Astronomy (San Francisco: W.H.
Freeman and Co.).
\bibitem{kn:Jetzer1} Ph. Jetzer, Astrophys. J. {\bf 432} (1994) L43
\bibitem{kn:Jetzer2}
Ph. Jetzer and E. Mass\'o, Phys. Lett. {\bf B 323} (1994) 347
\bibitem{kn:Masso} Ph. Jetzer and E. Mass\'o, 
in the proceedings of the second Rome workshop: ``The dark side of the
Universe: experimental efforts and theoretical frameworks'' 
(World Scientific, Singapore) (1996) 31
\bibitem{kn:je} Ph. Jetzer, Helv. Phys. Acta {\bf 69}, 179 (1996)
\bibitem{kn:Ingrosso} F. De Paolis, G. Ingrosso and Ph. Jetzer,
Astrophys. J. {\bf 470}, 493 (1996)
\bibitem{kn:Diploma} L. Grenacher, Diploma thesis University of
Z\"urich (1997)
\bibitem{de} F. De Paolis, G. Ingrosso, Ph. Jetzer and M. Roncadelli,
Phys. Rev Lett. {\bf 74}, 14 (1995)
\bibitem{de1} F. De Paolis, G. Ingrosso, Ph. Jetzer and M. Roncadelli,
Astron. and Astrophys. {\bf 295}, 567 (1995)
\bibitem{de2} F. De Paolis, G. Ingrosso, Ph. Jetzer and M. Roncadelli,
Comments on Astrophys. {\bf 18}, 87 (1995)
\bibitem{de3} F. De Paolis, G. Ingrosso, Ph. Jetzer and M. Roncadelli,
Astrophys. and Space Science {\bf 235}, 329 (1996)
\bibitem{de4} F. De Paolis, G. Ingrosso, Ph. Jetzer and M. Roncadelli,
Int. J. Mod. Phys. {\bf D5}, 151 (1996)
\bibitem{kn:Pfenniger}
D. Pfenniger, F. Combes and L. Martinet, Astron. and Astrophys. {\bf 285}
(1994) 79
\bibitem{kn:Silk1}
O.E. Gerhard and J. Silk, Astrophys. J. {\bf 472} (1996) 34
\bibitem{di} F. De Paolis, G. Ingrosso, Ph. Jetzer, A. Qadir and M. Roncadelli,
Astron. and Astrophys. {\bf 299}, 647 (1995)
\bibitem{kn:ber} V.S. Berezinsky, P. Blasi and V.S. Ptuskin,
astro-ph 9609048

\end{thebibliography}
\end{document}